\documentclass{ifacconf}

\usepackage{graphicx}      
\usepackage{natbib}        
\usepackage{siunitx}
\usepackage{pgfplots,filecontents}
\usepackage{tikz}
\usepackage{tikzscale}
\usepackage{amssymb}
\usepackage{amsmath}
\usepackage{algpseudocode}
\usepackage{algorithm}
\usepackage{fancyhdr}
\usepackage{caption} 
\usepackage{subcaption} 
\usepackage{eso-pic}

\graphicspath{{Figures_paper/}}

\definecolor{myorange}{cmyk}{0,0.35,0.85,0} 
\definecolor{mypurple}{cmyk}{0.5,1,0,0} 

\definecolor{matblue1}{rgb}{0,0.4470,0.7410}
\definecolor{matred1}{rgb}{0.85,0.325,0.098}
\definecolor{matyel1}{rgb}{0.9290, 0.6940, 0.1250}
\definecolor{matpur1}{rgb}{0.4940, 0.1840, 0.5560}
\definecolor{matgre1}{rgb}{0.4660, 0.6740, 0.1880}
\definecolor{matblue2}{rgb}{0.3010, 0.7450, 0.9330}
\definecolor{matred2}{rgb}{0.6350, 0.0780, 0.1840}
\definecolor{matgrey1}{rgb}{0.5, 0.6, 0.7}
\definecolor{matpink1}{rgb}{1, 0.07, 0.65}
\definecolor{matblue3}{rgb}{0.07, 0.62, 1}

\newcommand{\bluedashdot}{\raisebox{2pt}{\tikz{\draw[-,matblue1,densely dash dot,line width = 0.9pt](0,0) -- (3mm,0);}}}

\newcommand{\reddash}{\raisebox{2pt}{\tikz{\draw[-,matred1,dashed,line width = 0.9pt](0,0) -- (3mm,0);}}}
\newcommand{\bluedash}{\raisebox{2pt}{\tikz{\draw[-,matblue1,dashed,line width = 0.9pt](0,0) -- (3mm,0);}}}

\newcommand{\blackdash}{\raisebox{2pt}{\tikz{\draw[-,black,dashed,line width = 0.9pt](0,0) -- (3mm,0);}}}

\newcommand{\blackline}{\raisebox{2pt}{\tikz{\draw[-,black,solid,line width = 0.9pt](0,0) -- (3mm,0);}}}
\newcommand{\blueline}{\raisebox{2pt}{\tikz{\draw[-,matblue1,solid,line width = 0.9pt](0,0) -- (3mm,0);}}}
\newcommand{\redline}{\raisebox{2pt}{\tikz{\draw[-,matred1,solid,line width = 0.9pt](0,0) -- (3mm,0);}}}
\newcommand{\orangeline}{\raisebox{2pt}{\tikz{\draw[-,myorange,solid,line width = 0.9pt](0,0) -- (3mm,0);}}}
\newcommand{\purpleline}{\raisebox{2pt}{\tikz{\draw[-,mypurple,solid,line width = 0.9pt](0,0) -- (3mm,0);}}}

\newcommand{\bluedot}{\raisebox{.7pt}{\tikz{\draw[-,matblue1,solid,line width = 1pt](0,0) circle (.7mm);}}}
\newcommand{\reddot}{\raisebox{.7pt}{\tikz{\draw[-,matred1,solid,line width = 1pt](0,0) circle (.7mm);}}}
\newcommand{\yeldot}{\raisebox{.7pt}{\tikz{\draw[-,matyel1,solid,line width = 1pt](0,0) circle (.7mm);}}}
\newcommand{\purdot}{\raisebox{.7pt}{\tikz{\draw[-,matpur1,solid,line width = 1pt](0,0) circle (.7mm);}}}
\newcommand{\gredot}{\raisebox{.7pt}{\tikz{\draw[-,matgre1,solid,line width = 1pt](0,0) circle (.7mm);}}}
\newcommand{\bluedott}{\raisebox{.7pt}{\tikz{\draw[-,matblue2,solid,line width = 1pt](0,0) circle (.7mm);}}}

\newcommand{\yelline}{\raisebox{2pt}{\tikz{\draw[-,matyel1,solid,line width = 0.9pt](0,0) -- (3mm,0);}}}
\newcommand{\purline}{\raisebox{2pt}{\tikz{\draw[-,matpur1,solid,line width = 0.9pt](0,0) -- (3mm,0);}}}

\newcommand{\redlinet}{\raisebox{2pt}{\tikz{\draw[-,matred2,solid,line width = 0.9pt](0,0) -- (3mm,0);}}}

\newtheorem{assumption}[thm]{Assumption}
\theoremstyle{definition}

\begin{document}
\AddToShipoutPictureBG*{%
	\AtPageUpperLeft{%
		\setlength\unitlength{1in}%
		\hspace*{\dimexpr0.5\paperwidth\relax}
		\makebox(0,-2)[c]{
			\parbox{\paperwidth}{ \centering
				Leontine Aarnoudse, Commmutation-Angle Iterative Learning Control for Intermittent Data, \\
				In {\em 21st IFAC World Congress}, Berlin, Germany, 2020}}%
}}

\begin{frontmatter}

\title{Commutation-Angle Iterative Learning Control for Intermittent Data: Enhancing Piezo-Stepper Actuator Waveforms} 

\thanks[footnoteinfo]{This work is part of the research programme VIDI with project number 15698, which is (partly) financed by the Netherlands Organization for Scientific Research (NWO).}

\author[First]{Leontine Aarnoudse} 
\author[First]{Nard Strijbosch} 
\author[Second]{Edwin Verschueren}
\author[First]{Tom Oomen} 

\address[First]{Eindhoven University of Technology, Department of Mechanical Engineering, P.O. Box 513, 5600 MB Eindhoven (e-mail: l.i.m.aarnoudse@student.tue.nl, n.w.a.strijbosch@tue.nl, t.a.e.oomen@tue.nl)}
\address[Second]{Thermo Fisher Scientific, Eindhoven, The Netherlands}

\begin{abstract}                
Piezo-stepper actuators are used in many nanopositioning systems due to their high resolution, high stiffness, fast response, and the ability to position a mover over an infinite stroke by means of motion reminiscent of walking. The aim of this paper is to develop a control approach for attenuating disturbances that are caused by the walking motion and are therefore repeating in the commutation-angle domain. A new iterative learning control approach is developed for the commutation-angle domain, that addresses the iteration-varying and non-equidistant sampling that occurs when the piezo-stepper actuator is driven at varying drive frequencies by parameterizing the input and error signals. Experimental validation of the framework on a piezo-stepper actuator leads to significant performance improvements.
\end{abstract}

\begin{keyword}
Iterative Learning Control, Feedforward Control, Motion Control Systems, Piezo Actuators, Micromechantronic Systems
\end{keyword}

\end{frontmatter}

\section{Introduction} \label{sec:intro}

Many nanopositioning systems use piezo-stepper actuators to meet increasing requirements for high precision positioning that arise due to developments in the field of nanotechnology. Applications, such as nano-motion stages (\cite{Merry2011e}) and scanning probe microscopy (\cite{DenHeijer2014}), require the high resolution, high stiffness, and fast response of the piezoelectric elements as well as a large mover stroke which is provided by a motion that resembles walking. There are various ways to implement this walking motion, for example by walking drives (\cite{Shamoto1997a}) or bi-morph legs (\cite{Uzunovic2015a}). \\
\\
During the walking motion of a piezo-stepper actuator, engagement and release between the piezo elements and the mover can lead to repeating disturbances (\cite{DenHeijer2014,Strijbosch2019}). The piezo-stepper actuator is actuated using waveforms that describe the mapping from the commutation angle to the input voltage of the piezo elements. The disturbances are repeating with the period of these actuating waveforms and lead to a nonlinear relation between commutation angle and mover position, for which control typically is difficult.\\
\\
For varying velocities, the error profile caused by these disturbances is varying. In industrial implementations, piezo-stepper actuators are driven using varying drive frequencies with a constant sampling frequency in the temporal domain. Since the disturbances are repeating with the period of the actuating waveform, they are varying in the temporal domain for varying drive frequencies. In the commutation-angle domain, in which the waveforms are repeating, the disturbances are repeating. However, the sampling in the commutation-angle domain is varying and possibly non-equidistant.\\
\\
Learning control approaches such as iterative learning control (ILC) can compensate iteration-invariant disturbances perfectly, but they may amplify iteration-varying disturbances. In ILC, a feedforward input signal is modified based on preceding experiments that use the same reference, so that the tracking error is reduced over iterations (\cite{Bristow2006a}). Iteration-invariant disturbances are compensated perfectly, but typical ILC approaches amplify trial-varying disturbances significantly (\cite{Oomen2017}). Therefore, temporal domain ILC is not suited for a piezo-stepper actuator and a commutation-angle domain approach is needed instead. \\
\\
Existing approaches to ILC outside of the temporal domain depend strongly on assumptions regarding the sampling in the spatial domain. In \cite{Hoelzle2016b}, a 2D spatial ILC framework for micro-additive manufacturing is developed, in which the output of the system is measured at an iteration-invariant number of discrete points in space. In \cite{Kong2015a}, phase-indexed ILC is developed for a walking robot that behaves almost periodically, for which it is assumed that in the limit stable periodic behavior is obtained. In \cite{Strijbosch2019} $\alpha$-domain ILC is developed to reduce the repeating disturbances for a piezo-stepper actuator through waveform enhancement. It is assumed that continuous-time signals are available, which is not feasible in a digital setting. For $\alpha$-domain ILC, the iteration-varying non-equidistant sampling needs to be considered explicitly.
\\
\\
Existing approaches to learning control for iteration-varying and non-equidistant sampling do not consider the specific sampling situation of commutation-angle ILC. In \cite{Li2017}, iteration-varying sampling caused by incomplete trials is considered, where the sampling is equidistant and it is assumed that a constant sampling length is obtained in the limit. A framework for ILC with non-equidistant sampling is proposed in \cite{Strijbosch2019d}. In this approach the output is sampled at a high rate, which is assumed to be a multiple of the lower rate at which the input is sampled. Only part of the sampled output signal is used, resulting in iteration-varying non-equidistant sampling. For a piezo-stepper actuator, the input and output signals are sampled at the same iteration-varying rate, which in general cannot be related to a constant nominal sampling rate. \\
\\
Although important progress has been made to extend ILC outside of the temporal domain as well as consider iteration-varying sampling, at present ILC is not applicable to systems such as piezo-stepper actuators that involve both position domain disturbances and intermittent sampling. This paper aims to develop such a framework for commutation-angle iterative learning control, suitable for systems such as a piezo-stepper actuator. This leads to the following two contributions:
\begin{enumerate}
	\item A framework for commutation-angle domain ILC with iteration-varying and non-equidistant sampling, that uses basis functions to parameterize the input signal, is proposed. 
	\item The framework is implemented during walking experiments with a piezo-stepper actuator, resulting in significant performance improvements.
\end{enumerate} 
Preliminary results of the research presented here are reported in \cite{Strijbosch2019}, in which it is assumed that continuous-time signals are available, and in which learning is only applied during so-called `clamping' experiments and not during the actual walking motion of the piezo-stepper actuator.\\
\\
This paper is organized as follows. In Section \ref{sec:problem}, the functioning of a piezo-stepper actuator is explained and the problem formulation is given. In Section \ref{sec:framework}, the proposed framework for commutation-angle domain ILC is presented. In Section \ref{sec:exp_results}, the framework is experimentally validated using a piezo-stepper actuator. Conclusions are given in Section \ref{sec:conclusion}. Proofs will be published elsewhere.

\section{Problem formulation} \label{sec:problem}

\subsection{Piezo-stepper actuators}
\begin{figure}
	\centering
	\includegraphics[width=.8\linewidth]{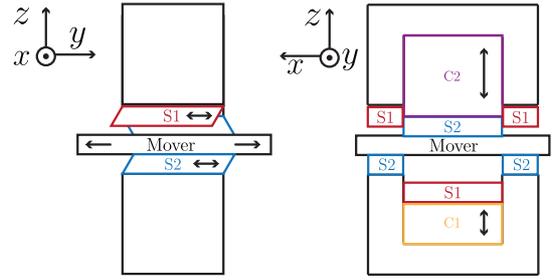}
	\caption{Schematic representation of a piezo-stepper actuator showing the clamp (`C') and shear (`S') elements of the first (\protect\redlinet, \protect\orangeline) and second (\protect\blueline, \protect\purpleline) group.}
	\label{fig:piezo_stepper}
\end{figure}

\begin{figure}
	\centering
	\includegraphics[width=.7\linewidth]{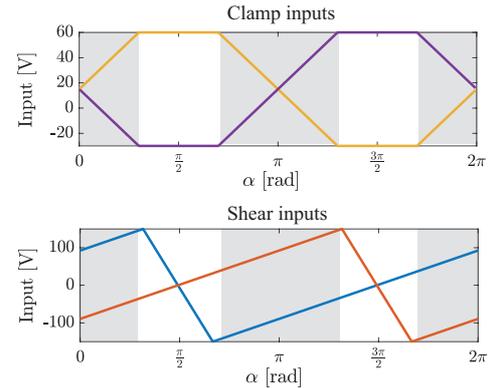}
	\caption{The waveforms for clamps 1 (\protect\yelline) and 2 (\protect\purline) contain regions where both clamps could be in contact with the mover, indicated in gray. In these regions the inputs for the shear elements 1 (\protect\redline) and 2 (\protect\blueline) have equal derivatives.}
	\label{fig:waveforms}
\end{figure}

Piezo-stepper actuators consist of a combination of longitudinal and shear elements in varying configurations, see e.g. \cite{Shamoto1997a,Uzunovic2015a}. The piezo-stepper actuator considered in this paper consists of two groups of piezo elements, each containing one longitudinal or `clamp' element and three shear elements, as shown in Fig. \ref{fig:piezo_stepper}. When the clamp element of a group is extended, the corresponding shear elements are in contact with the mover. The mover follows the displacement of the connected shear elements. Alternating the two piezo groups results in a walking motion, which leads to an unlimited stroke of the mover.\\
\\
The cyclic walking motion of the piezo-stepper is implemented using the periodic waveforms shown in Fig. \ref{fig:waveforms}. These waveforms map the commutation angle $\alpha \in [0,2\pi) \, [\si{\radian}]$ to the input voltage of the piezo elements. The number of steps per second is determined by the drive frequency $f_\alpha \, [\si{\hertz}]$. This drive frequency is integrated to obtain the commutation angle, i.e.,
\begin{equation}
\alpha (t) = 2\pi \int_0^t f_\alpha(\tau) \: \mathrm{d}\tau.
\label{eq:alpha_time}
\end{equation}
For a single step, it holds that $\alpha(0) = 0$ and $\alpha(T) = 2\pi$, where the duration $T \, [\si{\second}]$ of a step depends on the drive frequency. The system is driven with varying drive frequencies and sampled at a constant frequency $f_s \, [\si{\hertz}]$ in the temporal domain. The vector containing the $\alpha$-values at which a sample is taken for a single step is given by
\begin{equation}
\bar{\alpha} = 2\pi \begin{bmatrix}
\int_0^h f_\alpha (\tau) \textrm{d}\tau & \int_0^{2h} f_\alpha (\tau) \textrm{d}\tau & ... & \int_0^{Nh} f_\alpha (\tau) \textrm{d}\tau
\end{bmatrix}^\top
\label{eq:sampling}
\end{equation}
with sample interval $h = f_s^{-1} \, [\si{\second}]$. The number of samples within a step is given by $N=\lfloor T f_s \rfloor$.

\subsection{Modeling}

The piezo-stepper actuator is modeled as a gain with a lumped disturbance in the $\alpha$-domain. It is assumed that rate-dependent effects such as creep and hysteresis are negligible, since these can be compensated using a separate feedforward (\cite{Fleminga}, ch. 2,11). The modeled piezo-stepper actuator is described in the $\alpha$-domain without any significant time-domain dynamics.\\
\\
The displacement of a single shear element $y_i \, [\si{\meter}]$ is given by
\begin{equation}
y_i(\alpha(t)) = c u_{si}(\alpha(t)), \: i=1,2, 
\end{equation}
with positive piezo constant $c \, [\si{\meter \per \volt}]$ and shear inputs $u_{s1} \, [\si{\volt}]$ and $u_{s2} \, [\si{\volt}]$. \\
\\
The waveforms are designed to obtain a linear relation between the commutation angle $\alpha$ and the mover displacement $y$, see also \cite{Aarnoudse2020a}. Therefore, the waveforms of the shear elements are designed such that they have equal derivatives for any $\alpha$ where both clamps could be in contact with the mover, as shown in Fig. \ref{fig:waveforms}. The part of the step where one of the clamps is completely retracted is used to reset the corresponding shear elements. Because the connected shear elements are always moving with the same velocity, the combination of the two shear inputs $u_{s1}(\alpha)$ and $u_{s1}(\alpha)$ is written as a single input $u_s(\alpha)  \, [\si{\volt}]$, satisfying
\begin{equation}
\label{eq:division_shears}
\frac{\delta u_s(\alpha)}{\delta \alpha} = 
\begin{cases}
\frac{\delta u_{s1}(\alpha)}{\delta \alpha} &\text{if } \alpha \in [\frac{\pi}{3},\frac{2\pi}{3}] \\
\frac{\delta u_{s1}(\alpha)}{\delta \alpha} &\text{if } \alpha \in [\frac{4\pi}{3},\frac{5\pi}{3}] \\
\frac{\delta u_{s1}(\alpha)}{\delta \alpha} = \frac{\delta u_{s2}(\alpha)}{\delta \alpha} &\text{otherwise}. 
\end{cases}
\end{equation}
The corresponding desired mover position is given by
\begin{equation}
y_d(\alpha) = h_0 u_s(\alpha)
\end{equation}
with piezo constant $h_0 \, [\si{\meter \per \volt}]$. During open-loop experiments the desired linear relation between commutation angle and mover position is not obtained due to disturbances. These disturbances are assumed to be relatable to the commutation angle, and are modeled by a lumped disturbance $d_\alpha(\alpha)  \, [\si{\meter}]$. Therefore, the position of the mover with disturbances is described by 
\begin{equation}
y(\alpha(t)) = h_0 u_s(\alpha(t)) + d_\alpha(\alpha(t)).
\end{equation}
When a single step of the piezo-stepper actuator is considered, the system is written in terms of $\alpha$ as
\begin{align}
y(\alpha) = h_0 u_s(\alpha) + d_\alpha(\alpha), \quad \alpha \in [0,2\pi).
\label{eq:system_u}
\end{align}

\subsection{Repeatability of the disturbances}
\begin{figure}
	\centering
	\begin{subfigure}{.49\textwidth}
		\centering
		\includegraphics[width=.9\linewidth]{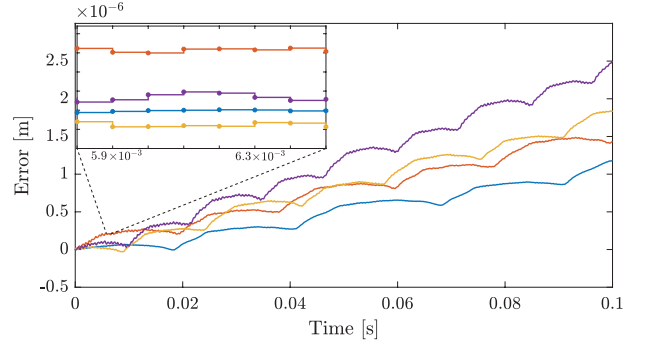}
		\caption{Disturbance as a function of time.}
		\label{fig:clamping_pos_time1}
	\end{subfigure}
	\\
	\begin{subfigure}{.49\textwidth}
		\centering
		\includegraphics[width=.9\linewidth]{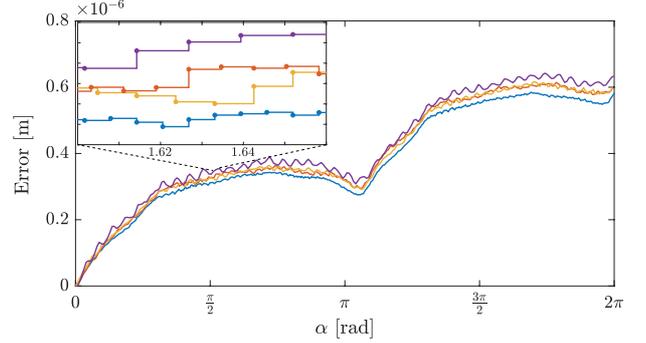}
		\caption{Disturbance as a function of $\alpha$.}
		\label{fig:clamping_pos_y1}
	\end{subfigure}
	\caption{Disturbances for a piezo-stepper during open-loop walking with drive frequencies \SI{20}{\hertz} (\protect\blueline), \SI{25}{\hertz} (\protect\redline), \SI{30}{\hertz} (\protect\yelline) and \SI{40}{\hertz} (\protect\purline). In the temporal domain (a) the sampling is equidistant (see zoom plot) but the disturbance is not repeating for different drive frequencies. In the $\alpha$-domain (b) the sampling is non-equidistant for varying drive frequencies, but the disturbances are similar.}
	\label{fig:clamping_OL_time1}
\end{figure}

During the walking motion, the desired linear relation between commutation angle and mover position is not obtained, since the position of the piezo-stepper actuator shows disturbances that are repeating with the period of the actuating waveform. These disturbances could be explained by physical sources, such as misalignment between the piezo elements. In Fig. \ref{fig:clamping_pos_time1} and \ref{fig:clamping_pos_y1}, the disturbance is plotted in the temporal domain and the $\alpha$-domain, respectively, for different drive frequencies. In the temporal domain the sampling is equidistant and equal for different drive frequencies, but the disturbance is drive-frequency dependent. In the $\alpha$-domain, the disturbance is repeating for different drive frequencies. However, the number of samples within a step and the distance between the samples in the $\alpha$-domain, given by (\ref{eq:sampling}), is varying for varying drive frequencies.\\ 
\\
The aim of this paper is to develop a framework for $\alpha$-domain iterative learning control with iteration-varying and non-equidistant sampling for systems with a dominant $\alpha$-domain repeating disturbance. This framework can be used to increase the performance and reduce the influence of the $\alpha$-domain repeating disturbance for a piezo-stepper actuator using the waveform enhancement method presented in \cite{Strijbosch2019}.

\section{$\alpha$-domain ILC with basis functions} \label{sec:framework}

In this section, the framework for $\alpha$-domain iterative learning control with basis functions is presented and convergence conditions are given.

\subsection{Assumptions}

The following assumptions are made for $\alpha$-domain ILC.
\begin{assumption}
	The initial condition $y_j(0)$ is identical for each iteration $j$.
	\label{as:init_cond}
\end{assumption}
\begin{assumption}
	The length of each iteration is constant in the $\alpha$-domain, i.e., $\alpha_j \in [0,2\pi)$.
	\label{as:trial_length}
\end{assumption}
\begin{assumption}
	For each iteration there is a unique mapping  $F_j : [0,T_j] \mapsto [0,2\pi)$ from the time interval $t \in [0,T_j]$ to the commutation-angle interval $\alpha \in [0,2\pi)$.
	\label{as:mapping}
\end{assumption}
\begin{assumption}
	The basis functions $\psi(\alpha)$ can be scaled to describe the $\alpha$-domain disturbance and reference exactly, i.e., $d_\alpha(\alpha) = \psi^\top(\alpha) \theta^d$ and $y_d(\alpha) = \psi^\top(\alpha) \theta^{yd}$.
	\label{as:perfect_fit}
\end{assumption}
Assumption \ref{as:init_cond} is satisfied by defining the initial position for each iteration to be $y(0)=0$. Assumption \ref{as:trial_length} is satisfied by choosing the measurement time so that $\alpha(0) = 0$ and $\alpha(T_j) = 2\pi$, which allows varying iteration lengths in the temporal domain. Assumption \ref{as:mapping} is satisfied when $\alpha$ is continuously increasing or decreasing within an iteration. Assumption \ref{as:perfect_fit} is satisfied by using suitable basis functions.

\subsection{Approach}
Iterative learning control in the $\alpha$-domain is used to compensate the $\alpha$-domain repeating disturbance at iteration-varying drive frequencies. ILC cannot be applied directly to the sampled input and output signals, since the sampling is iteration-varying. The vector containing the $\alpha$-values at which a sample is taken in iteration $j$ is given by $\bar{\alpha}_j \in \mathbb{R}^{N_j\times 1}$ according to (\ref{eq:sampling}). Since $\bar{\alpha}_j$ is iteration-varying, and the sample points may be non-equidistant, the input and error signals are parameterized using basis functions to obtain continuous descriptions. ILC is applied to these continuous signals, and the learned input signal is sampled for implementation. The continuous system to which ILC is applied is given by
\begin{align}
y_j(\alpha) &= h_0 (u_s(\alpha) + u_j(\alpha)) + d_\alpha (\alpha) \label{eq:y_j_cont}, \\
e_j(\alpha) &= y_d(\alpha) - y_j(\alpha) \label{eq:e_j_cont},
\end{align}
with standard shear input $u_s(\alpha) = \psi(\alpha)^\top \theta^{us}$, disturbance-compensating input $u_j(\alpha)$ and error $e_j(\alpha)$. The compensating input is constructed using basis functions as 
\begin{align}
u_{j+1}(\alpha) = \psi^\top(\alpha) \theta^u_{j+1} 
\label{eq:input}.
\end{align} 
The basis function vector $\psi$, containing $M$ linearly independent basis functions, and parameter vector $\theta^u_j$ are given by
\begin{align}
\psi(\alpha) &= 
\begin{bmatrix}
\psi_1(\alpha) & \psi_2(\alpha) & \dots & \psi_M(\alpha) 
\end{bmatrix}^\top \in \mathbb{R}^{M \times 1}, \\
\theta^u_j &= 
\begin{bmatrix}
\theta^u_{1,j} & \theta^u_{2,j} & \dots & \theta^u_{M,j}
\end{bmatrix}^\top \in \mathbb{R}^{M \times 1}.
\end{align}
An outline of the approach to $\alpha$-domain ILC for $n$ iterations is given in Algorithm \ref{alg:ILC_rbf}.

\begin{algorithm}[H]
	\caption{Approach to $\alpha$-domain ILC} 	\label{alg:ILC_rbf}
	\begin{algorithmic}	
		\item[]
		\State{Choose a basis $\psi$, see Section \ref{sec:exp_results}}

\State{\textbf{for} $j=1:n$}
		\State{\quad \: Perform an experiment for one step with $f_{\alpha,j}$}
		\State{\quad \: Find $\theta^e_j$ using a least squares fit as described in \\ \quad \: Section \ref{subsec:error}}
		\State{\quad \: Update the input parameters $\theta^u_{j+1}$ as described in \\ \quad \: Section \ref{subsec:ILC}}
		\State{\quad \: Update the input according to (\ref{eq:input}): $u_{j+1} = \psi^T \theta^u_{j+1}$}	
		\State{\quad \: Divide $u_{j+1}$ into waveforms $s_{1,j+1}$ and $s_{2,j+1}$ \\ \quad \: according to Section \ref{subsec:exp_results}}
\State{\textbf{end}}
	\end{algorithmic}
\end{algorithm}

\subsection{Error parameterization} \label{subsec:error}
The sampled error signal is parameterized using the same set of functions that forms the basis for the input $u_j$, since the system is assumed to behave as a gain. For each iteration, the error is sampled for all $\alpha \in \bar{\alpha}_j$, resulting in the sampled error signal $\bar{e}_j(\bar{\alpha}_j)$. Using (\ref{eq:y_j_cont}), (\ref{eq:e_j_cont}), (\ref{eq:input}) and Assumption \ref{as:perfect_fit}, the parameterized error for iteration $j$ is given by
\begin{align}
e_j(\alpha) = \psi^\top(\alpha) (\theta^{yd}  - h_0 (\theta^{us}+\theta_j) - \theta^d) = \psi^\top(\alpha) \theta_j^e.
\label{eq:error_psi2}
\end{align}
The vector of parameters $\theta^{e}_j \in \mathbb{R}^{M\times 1}$ that provides the optimal fit for $\bar{e}_j$ at the sample points $\bar{\alpha}_j$ is determined using a least squares fit, where the following cost function is minimized:
\begin{align}
\mathcal{J}_e(\theta^e_j) = \sum_{i=1}^{N_j} (\bar{e}_j(i) - \psi^\top(\bar{\alpha}_j(i)) \theta^e_j)^2 \label{eq:cost_e} 
\end{align} 
The following theorem gives the optimal parameter vector $\theta^e_j$ for the fit of the sampled error.
\begin{thm}
	\label{the:error_fit}
	Consider the sampled error vector $\bar{e}_j$ of the system described by (\ref{eq:y_j_cont}) and (\ref{eq:e_j_cont}) for which a fit over a continuous domain is given by $e_j = \psi^\top(\alpha) \theta^e_j$. If the sampled basis functions in $\bar{\psi}_{j}$ are linearly independent, the parameter vector that gives the unique least-squares optimal fit in terms of the cost function (\ref{eq:cost_e}) for $\bar{e}_j$ is given by:
	\begin{align}
	\theta^{e}_j &= (\bar{\psi}_{j} \bar{\psi}_{j}^\top)^{-1} \bar{\psi}_{j} \bar{e}_j, \text{ with} \\
	\bar{\psi}_j &= \begin{bmatrix}
	\psi_1(\bar{\alpha}_j) & \psi_2(\bar{\alpha}_j) & ... & \psi_M(\bar{\alpha}_j)
	\end{bmatrix}^\top .
	\end{align}
\end{thm}

\subsection{ILC update law} \label{subsec:ILC}
A continuous ILC update law is developed to determine the input parameters $\theta^u_{j+1}$ for iteration $j+1$, using the continuously defined input and error signals of iteration $j$. To find the optimal input parameters, the continuous cost function $\mathcal{J}$ is minimized, which is given by
\begin{align}
\mathcal{J}(\theta^u_{j+1}) = & \int_{0}^{2\pi} \left( W_e(\alpha) e_{j+1}(\alpha)^2  
+   W_u(\alpha) u_{j+1}(\alpha)^2 \right. \nonumber \\  \label{eq:cost}
&  \left. +   W_{\Delta u}(\alpha) (u_{j+1}(\alpha)-u_j(\alpha))^2 \right) \: \mathrm{d} \alpha. 
\end{align}
The weights $W_e(\alpha)$, $W_u(\alpha)$ and $W_{\Delta u}(\alpha)$ are non-negative functions that are tuned to obtain certain performance and robustness properties. $W_e(\alpha)$ influences the performance of the learning, $W_u(\alpha)$ influences the robustness against model uncertainty and $W_{\Delta u}(\alpha)$ influences the attenuation of iteration-varying disturbances.\\ 
\\
\begingroup
\allowdisplaybreaks
The following theorem gives the optimal update for the vector of input parameters $\theta^u_j$.
\begin{thm}
	\label{the:update_rbf_ilc}
	Consider the system described by (\ref{eq:y_j_cont}) and (\ref{eq:e_j_cont}) for which the input $u_{j+1}(\alpha) = \psi^\top(\alpha)\theta^u_{j+1}$ of iteration $j+1$ is constructed using any given set of linearly independent basis functions $\psi(\alpha)$. The cost function (\ref{eq:cost}) with non-negative weight functions $W_e(\alpha)$, $W_u(\alpha)$ and $W_{\Delta u}(\alpha)$, of which at least one is positive for all $\alpha \in [0,2\pi)$, leads to an optimal update of the parameters $\theta^u_{j+1}$ given by 
	\begin{align}
	\theta^u_{j+1} &= Q_\psi \theta^u_j + L_\psi \theta^e_j, \quad \text{with} \label{eq:update_rbf} \\ \nonumber 
	Q_\psi &= \Big(  \int_{0}^{2\pi} \left(h_0^2 W_e(\alpha) + W_u(\alpha)  \right.  \\ \nonumber 
	& \left.  + W_{\Delta u}(\alpha)\right) \psi(\alpha) \psi^\top(\alpha) \: \mathrm{d}\alpha \Big)^{-1} \\ \nonumber  
	&\int_{0}^{2\pi} \left(h_0^2 W_e(\alpha) + W_{\Delta u}(\alpha)\right) \psi(\alpha) \psi^\top(\alpha) \: \mathrm{d}\alpha \\ \nonumber & \\ \nonumber
	L_\psi &= \Big(  \int_{0}^{2\pi} \left(h_0^2 W_e(\alpha) + W_u(\alpha)  \right.  \\ \nonumber 
	& \left. + W_{\Delta u}(\alpha)\right) \psi(\alpha) \psi^\top(\alpha) \: \mathrm{d}\alpha \Big)^{-1} \\ \nonumber & \int_{0}^{2\pi}  W_e(\alpha) h_0 \psi(\alpha) \psi^\top(\alpha) \: \mathrm{d}\alpha.
	\end{align}
\end{thm}
\endgroup

\subsection{Monotonic convergence} \label{subsec:conv}
To avoid large learning transients and ensure convergence to a unique input signal, the ILC system needs to be monotonically convergent. Conditions for monotonic convergence of $\alpha$-domain ILC with basis functions are given by the following theorem.
\begin{thm}
	\label{the:conv_rbf_ilc}
	Consider the system described in (\ref{eq:y_j_cont}) and (\ref{eq:e_j_cont}) with input (\ref{eq:input}). For the update law (\ref{eq:update_rbf}), with non-negative weight functions $W_e(\alpha)$, $W_u(\alpha)$ and $W_{\Delta,u}(\alpha)$ of which at least $W_e(\alpha)$ or $W_u(\alpha)$ is positive for all $\alpha \in [0,2\pi)$, and linearly independent basis functions $\psi$, the sequence of parameter vectors $\{\theta^u_j\}_{j\in \mathbb{N}}$ is monotonically convergent in the 2-norm towards a fixed parameter vector $\theta^{u*}$. 
\end{thm}

\section{Experimental results}		\label{sec:exp_results}
To validate the framework for $\alpha$-domain ILC with basis functions, it is applied during a series of walking experiments with a piezo-stepper actuator. First, suitable basis functions are selected. Then, experimental results are presented.

\subsection{Basis function selection} \label{subsec:basis}
\begin{figure}
	\centering
	\includegraphics[width=.9\linewidth]{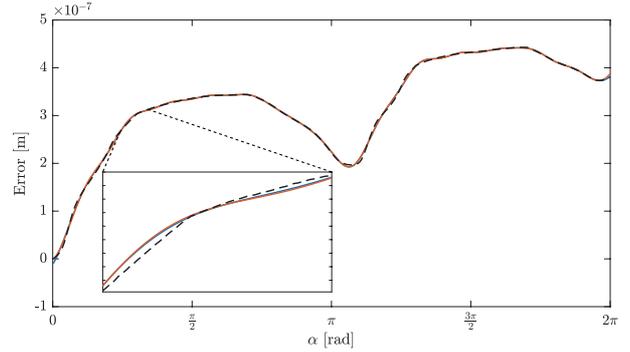}
	\caption{Comparison between the sampled error signal (\protect\blackdash) and fits using 30 inverse quadratic radial basis functions and 1000 (\protect\blueline) or 50 (\protect\redline) samples.}
	\label{fig:fit_equidistant_30}
\end{figure}

The input and error signals are parameterized using a set of 30 inverse quadratic radial basis functions. Basis functions in ILC are typically chosen based on prior knowledge regarding the disturbance or reference, such as the origin (\cite{Bolder2014}) or the shape (\cite{Mishra2009}). In $\alpha$-domain ILC, the basis functions are chosen based on how well they describe the sampled error signal. The $M$ inverse quadratic radial basis functions are linearly independent (\cite{Schaback2006}, ch. 5) and given by
\begin{align}
\psi_k(\alpha) = \frac{1}{1+(\|\alpha-c_k\|)^2}, \quad k = 1,2,...,M,
\end{align}
where the center points $c_k$ of the radial basis functions are divided equidistantly over the domain $[0,2\pi)$. Note that there are other valid choices for basis functions.\\
\\
The set of 30 inverse quadratic radial basis functions is used to fit error signals with 1000 and 50 equidistant samples, as shown in Fig. \ref{fig:fit_equidistant_30}. For a fit using 1000 samples, the root mean square (RMS) of the difference between sampled error signal and fit is approximately $2.4\times 10^{-9}$. The RMS value of the difference between this fit and a fit using a downsampled signal of 50 samples is approximately $8.6 \times 10^{-10}$. Therefore, it is concluded that the basis functions describe the error well and that the influence of the number of available sample points on the quality of the fit is negligible, provided that the number of samples is larger than or equal to the number of basis functions. 

\subsection{Experimental results} \label{subsec:exp_results}
\begin{figure}
	\centering
	\includegraphics[width=.7\linewidth]{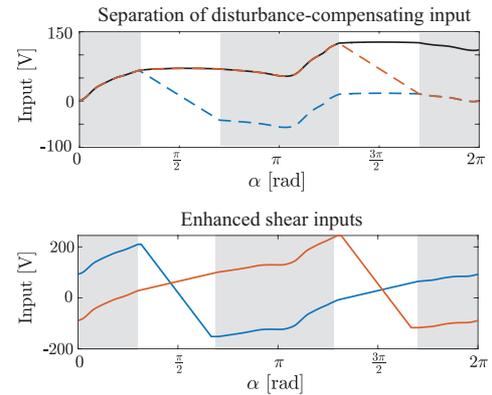}
	\caption{Waveform enhancement using a learned disturbance-compensating input signal. Regions where both groups could be in contact with the mover are indicated in gray. A compensating input signal (\protect\blackline) is divided into two inputs for the shear elements 1 (\protect\reddash) and 2 (\protect\bluedash). These inputs are added to the standard waveforms, resulting in enhanced waveforms for shear elements 1 (\protect\redline) and 2 (\protect\blueline).}
	\label{fig:waveform_enhancement}
\end{figure}

The input signal $u_j(\alpha)$, which is learned using ILC, is separated into two inputs $u_{1,j}(\alpha)$ and $u_{2,j}(\alpha)$ for the two groups of shear elements. The inputs satisfy (\ref{eq:division_shears}), such that when both shear elements are in contact with the mover their velocities are identical. These shear inputs are added to the standard waveforms, resulting in enhanced waveforms as is shown in Fig. \ref{fig:waveform_enhancement}. 

\begin{figure}
	\centering 
	\includegraphics[width=.9\linewidth]{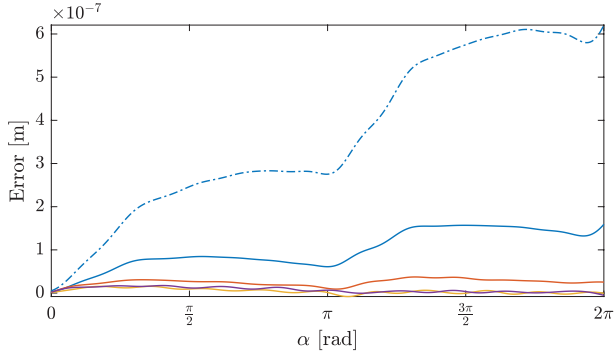}
	\caption{Error signal during a step for iterations 0 (\SI{30}{\hertz}, \protect\bluedashdot), 3 (\SI{30}{\hertz}, \protect\blueline), 7 (\SI{35}{\hertz}, \protect\redline), 11 (\SI{25}{\hertz}, \protect\yelline) and 15 (\SI{28}{\hertz}, \protect\purline). Between iterations 0 and 15 the RMS value of the error is reduced by a factor 35.}
	\label{fig:rbf_high_comp2}
\end{figure}

\begin{figure}
	\centering
	\includegraphics[width=.9\linewidth]{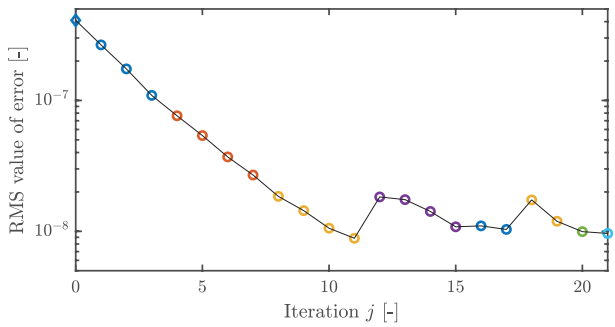}
	\caption{Convergence of the RMS value of the error during an open-loop walking experiment with $W_e = 1$, $W_u = 0$ and $W_{\Delta u} = 4.7\times 10^{-17}$. Subsequent drive frequencies: \SI{30}{\hertz}(\protect\bluedot), \SI{35}{\hertz}(\protect\reddot), \SI{25}{\hertz}(\protect\yeldot), \SI{28}{\hertz}(\protect\purdot), \SI{22}{\hertz}(\protect\gredot), \SI{20}{\hertz}(\protect\bluedott).}
	\label{fig:rbf_high_conv2}
\end{figure}

\begin{figure}
	\centering
	\includegraphics[width=.9\linewidth]{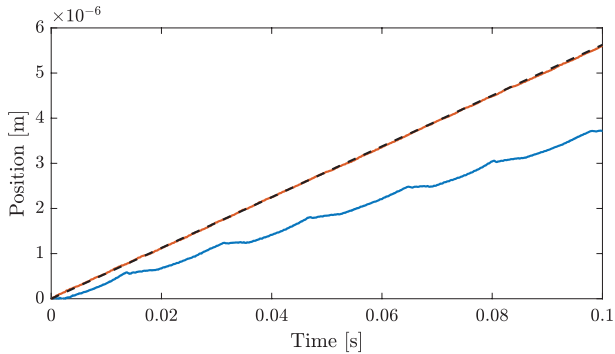}
	\caption{The position of the mover with standard shear waveforms (\protect\blueline)  and $f_\alpha = 30 \, \si{\hertz}$ deviates from the reference (\protect\blackdash). The enhanced shear waveforms of iteration 17 compensate the disturbance such that the position (\protect\redline) approximates the reference. }
	\label{fig:pos_improv_high_all}
\end{figure}

In the experiments scalar cost function weights are used, i.e., $W_e(\alpha) = W_e \, \forall \, \alpha \in [0,2\pi)$ etc.. The desired linear relation between the commutation angle $\alpha$ and the mover position is described by the reference $y_d(\alpha) = 3\times 10^{-7} \alpha.$ \\
\\
During an open-loop walking experiment with iteration-varying drive frequencies ranging between $20-35 \, \si{\hertz}$, the error is reduced significantly over iterations, as shown in Fig. \ref{fig:rbf_high_comp2}. The RMS value of the continuous error signal $e_j(\alpha)$, given by $RMS(e_j) = \sqrt{\frac{1}{2\pi} \int_0^{2\pi} e_j(\alpha)^2 \mathrm{d}\alpha}$, is shown in Fig. \ref{fig:rbf_high_conv2} and converges to a bounded region. At iterations 12 and 18, a change in drive frequency causes an increase of the RMS value of the error. This is caused by rate-dependent behaviors in the piezo shear elements, which can be compensated using feedforward (\cite{Croft2001}).\\
\\
The improvements in the temporal domain are shown in Fig. \ref{fig:pos_improv_high_all}, where the position of the mover for iterations 0 and 17 is compared. It is shown that the repeating disturbance is compensated, so that a linear relation between commutation angle and mover position is obtained.

\section{Conclusion}	\label{sec:conclusion}

A new framework for $\alpha$-domain iterative learning control is presented that is capable of fully mitigating repeatable disturbances in the $\alpha$-domain for a piezo-stepper actuator, while coping with iteration-varying and non-equidistant measurement and actuation points. Basis functions are used to parameterize the input and error signals and obtain continuous descriptions. These continuous descriptions are used in an optimal ILC update law. Compensation of the $\alpha$-domain repeating disturbances for a piezo-stepper actuator during walking experiments results in a linear relation between commutation angle and mover position. This improves the positioning accuracy and reduces the complexity of closed-loop control in an industrial setting.

\bibliography{C:/Users/s143007/Documents/MendeleyBIB/library}


\end{document}